\documentclass{emulateapj}
\usepackage{bm}
\usepackage{textcomp}
\usepackage{natbib}
\usepackage[colorlinks,linkcolor=blue,
urlcolor=blue,citecolor=blue,linktocpage=true,plainpages=false,breaklinks]{hyperref}
\shortauthors{Smitha et al.}

\begin{document}
\title{The quantum interference effects in the Sc {\sc ii} 4247~\AA\ line 
of the Second Solar Spectrum}
\author{H. N. Smitha$^{1}$, K. N. Nagendra$^{1}$, J. O. Stenflo$^{2,3}$, 
M. Bianda$^{3}$,
and R. Ramelli$^{3}$} 
\affil{$^1$Indian Institute of Astrophysics, Koramangala, Bangalore, India}
\affil{$^2$Institute of Astronomy, ETH Zurich, CH-8093 \  Zurich, Switzerland }
\affil{$^3$Istituto Ricerche Solari Locarno, Via Patocchi, 6605 Locarno-Monti, Switzerland}
\shortauthors{Smitha et al.}
\email{smithahn@iiap.res.in, knn@iiap.res.in, stenflo@astro.phys.ethz.ch, 
mbianda@irsol.ch, ramelli@irsol.ch}
\begin{abstract}
The Sc~{\sc ii} 4247\,\AA\ line formed in the chromosphere is one of the lines 
well known,
like the Na~{\sc i} D$_2$ and Ba~{\sc ii} D$_2$, for its prominent triple peak 
structure in $Q/I$ 
and the underlying quantum interference effects governing it. In this paper,
we try to study the nature of this triple peak structure using the theory of 
$F$-state interference including the effects of partial frequency redistribution (PRD)
and radiative transfer (RT). We compare our results with the observations taken 
in a quiet region near the solar limb. In spite of accounting 
for PRD and RT effects it has not been possible to reproduce the observed triple peak 
structure in $Q/I$. While the two wing PRD peaks (on either side of central peak)
and the near wing continuum can be reproduced, the central peak is 
completely suppressed by the enhanced depolarization resulting 
from  the hyperfine structure splitting. 
This suppression remains for all the tested widely different 1D model 
atmospheres or for any multi-component combinations of them. While 
multidimensional radiative transfer effects may improve the fit to 
the intensity profiles, they do not appear capable of explaining the 
enigmatic central $Q/I$ peak. This leads us to suspect that some aspect 
of quantum physics is missing.
\end{abstract}

\maketitle

\section{Introduction}
\label{intro}
With the advent of highly sensitive spectropolarimeters like the 
Zurich Imaging Polarimeter (ZIMPOL) we now have access to the linearly 
polarized spectrum of the Sun that is due to coherent scattering processes 
in the Sun's atmosphere (and which has nothing to do with the well-known 
transverse Zeeman effect). This new linearly polarized spectrum of the 
Sun is commonly referred to as the ``Second Solar Spectrum" 
\citep{1996Natur.382..588S,1997A&A...321..927S}. It is richly 
structured with signatures of different kinds of scattering processes 
taking place in atomic systems of varying complexity. Of particular 
interest are the many often enigmatic signatures of quantum interference 
effects between fine structure states, hyperfine structure states, and 
magnetic substates (Hanle effect).

Atoms with non-zero electron spin $S$ undergo fine structure splitting 
and exhibit $J$-state interference whereas the atoms with non-zero nuclear spin $I_s$
undergo hyperfine structure splitting (HFS) and show $F$-state interference. 
The Sc~{\sc ii} 4247\,\AA\ line is governed by $F$-state interference. 

\begin{figure}[ht]
\centering
\includegraphics[width=5.0cm]{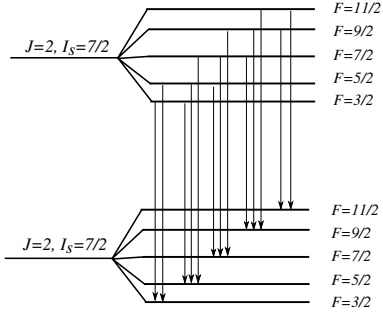}
\caption{Level diagram showing the hyperfine structure splitting 
of the $3d4s$ and $3d4p$ atomic levels of the Sc~{\sc ii} atom.}
\label{level-diag}
\end{figure}

Here we extend our previous work on the Ba~{\sc ii} D$_2$ line 
\citep{2013ApJ...768..163S},
to study the Sc~{\sc ii} line at 4247~\AA. This line arises due to the transition $J=2 \to J=2$.
Due to {coupling with} the nuclear spin ($I_s=7/2$) 
both the upper and the lower $J$ levels are split into five $F$-states each with 
13 radiative transitions between them. The level diagram of this system 
is shown in Figure~{\ref{level-diag}}. 
We use the theory of $F$-state interference 
presented in \citet[][see also \citealt{2012ApJ...758..112S}]{2013ApJ...768..163S},
which takes account of the partial frequency redistribution (PRD) effects
in the absence of magnetic fields. 
The results in the present paper
do not include the contributions from magnetic fields.
The theory of $F$-state interference in the 
presence of magnetic fields including the effects of PRD 
has been recently developed in \cite{2014ApJ...786..150S}.

The 4247\,\AA\ line of Sc~{\sc ii} is a chromospheric line with an approximate 
height of formation between 900-1100~km above the photosphere. $^{45}$Sc is the only
stable isotope of scandium. It shows prominent triple peak structure in its 
$Q/I$ spectra \citep[see][]{2002sss..book.....G,2003ASPC..307..385S}.
Modeling of this triple peak structure  using the last scattering approximation 
was attempted by \cite{2009A&A...508..933B}. 
The effects of PRD and radiative transfer were neglected in that work. 

In the present paper, by taking account of both PRD and radiative transfer effects, 
we study the sensitivity of the $(I,Q/I)$ profiles to different atomic and atmospheric
parameters. From our efforts we find it difficult to reproduce the triple peak 
structure in $Q/I$ and also the rest intensity. 

The rest intensity is dependent on the model atmospheric properties.
A plausible reason for the failure of reproducing the rest intensity is the use of 
1D model atmospheres.  
In several earlier works such as \cite{2007A&A...467..695H,2011A&A...529A.139S,
2013ApJ...768..163S,2014arXiv1407.5461S}, difficulties with the 1D models were  encountered.
The limitations of the 1D model atmospheres are also described in 
\cite{1997fsp..proc..217K, 2002JAD.....8....8R, 2011ApJ...736...69U}.
We believe that the 3D model atmospheres may improve the fit to the 
observed rest intensity in case of the Sc~{\sc ii}~4247~\AA\ line.

In $Q/I$, the central peak is 
suppressed due to depolarization from HFS. However, the PRD peaks and the near wing 
continuum in the theoretical profiles match closely with the observed profiles. 
Our tests suggest that the observed $Q/I$ profiles cannot be 
reproduced by modifications in the existing 1D standard model atmospheres. 
Hence we suspect the role of other physical effects in 
shaping the observed profiles, which may not have been accounted for in the
 present treatment. 
The lower level Hanle effect could qualify as being one such
effect which can increase the polarization of the central peak,
but its contribution is significant only for fields $\le1$G 
\citep{2009A&A...508..933B}.

The details of the various tests conducted by us are discussed in the sections below.

\section{Computing the theoretical profiles}
The details of computing the Stokes profiles with $F$-state interference including the effects of 
PRD and radiative transfer using realistic 1D model atmospheres are 
presented in \cite{2013ApJ...768..163S}. 
We use the same here too and hence do not repeat them. 
However, certain physical quantities need to be redefined to 
represent the Sc~{\sc ii} 4247\,\AA\ line system, and they are presented below.\\

\noindent{\it The Voigt profile function:} For the case of
Sc~{\sc ii} 4247\,\AA\ line, the Voigt profile function defined in Equation~(3) of \cite{2013ApJ...768..163S}
 is to be replaced by
\begin{eqnarray}
&&\phi(\lambda,z)= \bigg[\frac{1}{25}\phi(\lambda_{3\,3},z)+ 
\frac{3}{50}\phi(\lambda_{3\,5},z) +\frac{3}{50}\phi(\lambda_{5\,3},z)  \nonumber \\ &&+
\frac{1}{1400}\phi(\lambda_{5\,5},z + \frac{5}{56}\phi(\lambda_{5\,7},z)+
\frac{5}{56}\phi(\lambda_{7\,5},z)\nonumber \\ && + \frac{2}{105}\phi(\lambda_{7\,7},z)+
\frac{11}{120}\phi(\lambda_{7\,9},z + \frac{11}{120}\phi(\lambda_{9\,7},z)  \nonumber \\ && +
\frac{25}{264}\phi(\lambda_{9\,9},z) + \frac{7}{110}\phi(\lambda_{9\,11},z)+
\frac{7}{110}\phi(\lambda_{11\,9},z)\nonumber \\ && + 
\frac{13}{55}\phi(\lambda_{11\,11},z)\bigg],
\label{comb-prof}
\end{eqnarray}
where $\phi(\lambda_{F_a\,F_b},z)$ is the Voigt profile function for the 
$F_a \to F_b$ transition with $F_a$ and $F_b$  being the 
initial and excited $F$-states respectively. For notational brevity, the subscripts $F_a$ and $F_b$
in the $\phi$ terms are multiplied by 2 in the above equation. \\

\noindent{\it The depolarizing elastic collision rate $D^{(2)}$:}  The branching ratios
which describe the contribution from type-II and collisional redistribution (type-III) are defined in 
Equation~(7) of \cite{2013ApJ...768..163S}. The depolarizing elastic collision rate $D^{(2)}$
which enter through the branching ratio $B^{(2)}$ can be computed 
using Equation~(7.102) of \cite{2004ASSL..307.....L}
\begin{equation}
 D^{(K)}(J) = C^{(0)}_E(J) - C^{(K)}_E(J),
\end{equation}
where $ C^{(K)}_E(J)$ is given by
\begin{eqnarray}
  C^{(K)}_E(J) = (-1)^K 
\frac{\left\lbrace 
\begin{array}{ccc}
J & J & K\\
J & J & \tilde{K}\\
\end{array}
\right\rbrace}
{\left\lbrace 
\begin{array}{ccc}
J & J & 0\\
J & J & \tilde{K}\\
\end{array}
\right\rbrace}
 C^{(0)}_E(J),
\end{eqnarray}
with $C^{(0)}_E(J)=\Gamma_E/(2J+1)$.

If the interaction between the atom and the colliding particle is
assumed to be of dipolar type then $\tilde{K}=1$. In this case
$D^{(2)}(J)=0.1\, \Gamma_E$. If the interaction is assumed to be 
dipole-dipole in nature, then $\tilde{K}=2$. In this case
$D^{(2)}(J)=0.243\, \Gamma_E$. 
We have tested that both these values of $D^{(2)}$ give nearly identical 
emergent $Q/I$ profiles. 
\begin{figure}[ht]
\centering
\includegraphics[width=6.8cm]{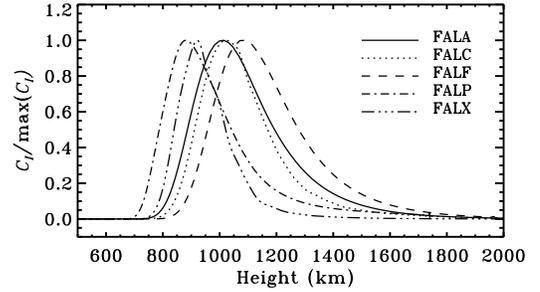}
\caption{Contribution function $C_I$ computed from the FAL model atmospheres
at $\mu=0.1$, for the line center wavelength.}
\label{contrib}
\end{figure}
\begin{figure}[ht]
\centering
\includegraphics[width=7.5cm]{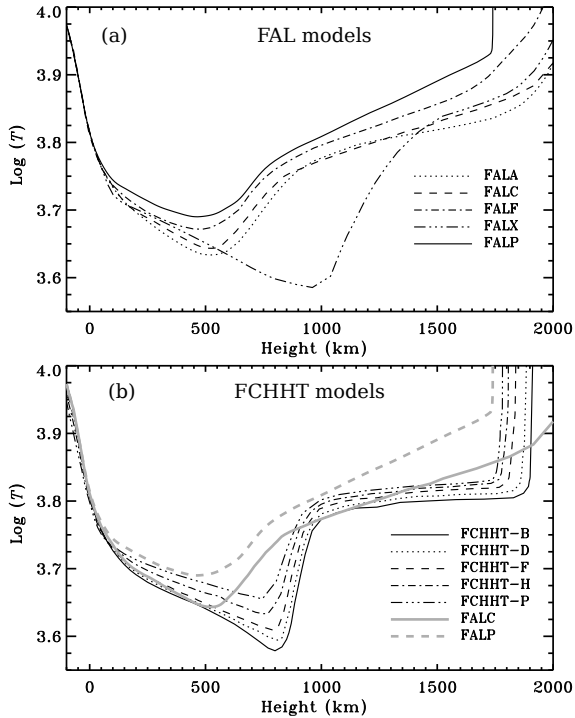}
\caption{Temperature structures of the FAL and FCHHT model atmospheres.}
\label{temp-struc}
\end{figure}
This is because the Sc~{\sc ii}~4247\,\AA\ line is formed at a height of 900-1100 km 
above the photosphere. This can be seen from Figure~\ref{contrib} where the 
intensity contribution functions $C_I$ are plotted as a function of height using
different FAL model atmospheres at line center wavelength. 
The temperature structures of the FAL models as a function of height is shown 
in Figure~\ref{temp-struc}a (see Section~\ref{fal-models} for a discussion on these models).

The function $C_I$ is defined as
\citep[see][]{1994ASSL..189.....S,1999A&A...341..902F}
\begin{equation}
C_{I}(\tau,\mu) = \frac{1}{\mu}S_{I}(\tau,\mu)\,e^{-\tau/\mu},
\label{c-i}
\end{equation}
with $d\tau=-k_{l}dz$, $k_l$ is the line absorption coefficient, and
\begin{equation}
 I_{\lambda}(\mu,z=\infty) = \int_{-\infty}^{\infty} C_{I}(z^\prime,\mu) dz'.
\end{equation}
From Equation~(\ref{c-i}), the contribution function is proportional to the 
source function $S_I$. Since $S_I$  differs for each model atmosphere, 
the contribution functions also differ. However, the maximum contributions from the FAL 
model atmospheres fall within the height range 900-1100km.

\begin{figure}[ht]
\centering
\includegraphics[width=6.8cm]{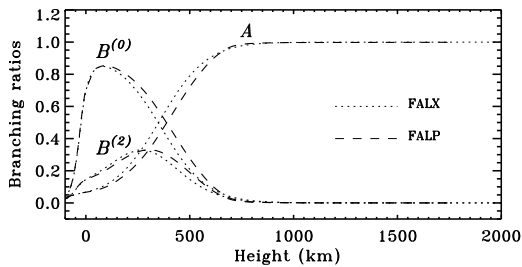}
\caption{ The branching ratios $A$ and $B^{(K)}$, with $K=0$ and $2$, 
as a function of height computed using the FALX and FALP model atmospheres.}
\label{branch}
\end{figure}

At these heights, the branching ratio
$B^{(2)}$ is nearly zero as seen from Figure~{\ref{branch}}.
We choose $D^{(2)}(J)=0.243\,\Gamma_E$ for further computations.

\section{Observations}
\label{obs}
The observations of the Sc~{\sc ii} 4247\,\AA\  line analyzed in this 
paper were recorded on September 15, 2012 at IRSOL, Switzerland using the
ZIMPOL-3 imaging polarimeter \citep{2010SPIE.7735E..66R}.
The photoelastic modulator (PEM) followed by a linear polarizer (beam splitter)
was used as the polarization analyzer. 

Though the telescope is almost free from instrumental polarization and cross talk effects around 
the equinox, to minimize residual instrumental signatures, a glass compensation 
plate was inserted in the optical path between the calibration optics and the analyzer. 
This also reduces the residual linear polarization offset. 
The optics was adjusted such that the positive $Q$ represents the 
linear polarization parallel to the spectrograph slit. 
An image derotator (Dove prism) placed between the analyzer and the slit-jaw 
allowed to rotate the solar image, and compensate for the solar rotation. The 
analyzer and the calibration optics were also rotated correspondingly.
The observations were performed at a quiet region with the spectrograph slit 
placed parallel to the solar East limb. The spectrograph grating angle 
and a prefilter were selected to work with the 13$^{\rm th}$ spectral order.
On the CCD we got a resolution of 1.44\arcsec\ per pixel along the 
spatial direction and 5.25 m\AA\ per pixel along the spectral direction.
Three measurements were obtained by placing the slit at 5\arcsec, 15 \arcsec, and 25\arcsec\ from the solar limb.
The observations at each $\mu$-position consisted of a sum of 1000 frames obtained with an exposure of 1 sec, 
making the total exposure time as 16 minutes.

\begin{figure}[htbp]
\centering
 \includegraphics[width=7.0cm]{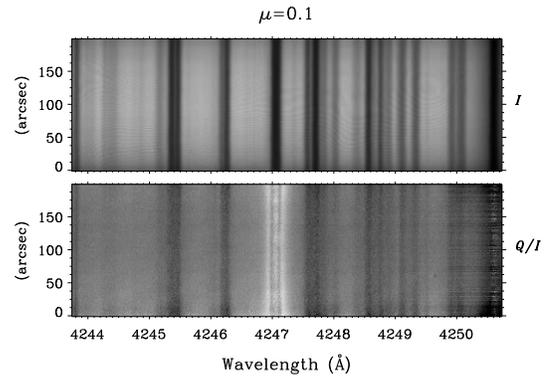}
\caption{CCD image showing $(I,Q/I)$ for the Sc~{\sc ii} 4247\,\AA\ line
recorded on September 15, 2012 using the ZIMPOL spectropolarimeter at IRSOL, Switzerland.}
\label{ccd-image}
\end{figure}
 
The image motion perpendicular to the limb was compensated with a glass tilt-plate. 
The tilt of the plate was determined automatically with a limb recognition software using
the information in the slit jaw image.
The Stokes $(I,Q/I)$ images shown in Figure \ref{ccd-image} were obtained after the data reduction.
We also did a flat-field recording by moving the telescope around 
the disk center while recording 20 frames. The flat-field 
observations were used to correct the intensity images. 
The observed $(I,Q/I)$ profiles used in this paper were obtained after performing a
spatial averaging from 60\arcsec\ to 140\arcsec\ along the slit.

\subsection{Determining the absolute zero level of  polarization}
\label{zero-level}

\begin{figure}[ht]
\centering
\includegraphics[width=6.8cm]{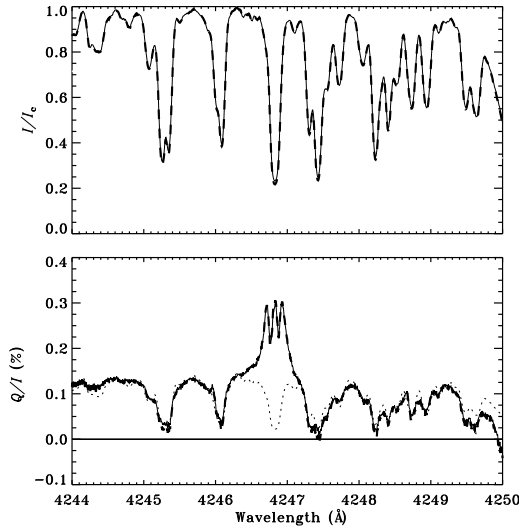}
\caption{Fit to the blend lines around the Sc~{\sc ii}~4247~\AA\ line profile
using Equation~(\ref{pc-ic}). The solid lines represent the observations, and dotted line 
in the second panel represents $p_c[1-(1-\frac{I}{I_c})^{0.6}] - p_0$. 
To obtain a fit we choose $p_c=0.15\%$ and $p_0=0.07\%$. The dashed lines show the 
observed profiles corrected for 2\% stray light. The dashed and solid lines nearly overlap.}
\label{sc-stray}
\end{figure}

The absolute zero level of polarization is determined 
using the blend lines as described in 
\citet[][see also \citealt{1998A&A...329..319S}]{2005A&A...429..713S}.
According to this method, the relative line depths of the depolarizing blend lines in
Stokes $I$ and $Q/I$ are related with the following one-parameter model as
\begin{equation}
 \left(\frac{p_c-p}{p_c}\right) = \left(\frac{I_c-I}{I_c}\right)^\alpha,
\label{pc-ic}
\end{equation}
where $I_c,\,p_c$ are the intensity and polarization of the continuum, 
and $I,\,p$ are the respective quantities for the blend lines. $\alpha$
is a free model parameter that determines the shape of the depolarizing lines. 
We choose $\alpha=0.6$ for further analysis.
Figure~\ref{sc-stray} shows the comparison between the observed profile (solid line)
and the profile computed using Equation~(\ref{pc-ic}, second panel: dotted line).
This dotted line represents $p_c[1-(1-\frac{I}{I_c})^{0.6}] - p_0$. 
Here $p_0$ is a free model parameter that represents the apparent level of the true zero point 
of the polarization scale. 
The blend line depth is sensitive to the 
value of $p_c$. To get the observed line depths we need $p_c=0.15\%$ $(p_{c,obs})$.
Also to match the solid and the dotted profiles a shift of $p_0=0.07\%$
has to be applied.
As seen from this figure, we obtain a good match between the 
solid and the dotted profiles for this set of parameters.

\subsection{Stray light correction}
\label{stray-correction}
Next, we applied a stray light correction of 2\% of the 
continuum to both $I$ and $Q/I$. For correcting the $Q/I$ profile
we have used the value of $p_c$ determined above.
The details of the steps followed are 
given in \cite{2014arXiv1407.5461S}.
The comparison between the observed profiles with stray light correction (dashed line) 
and without (solid line) is shown in Figure~{\ref{sc-stray}}. 
The stray light corrected observed profiles 
nearly overlap with the profiles without this correction. 

\section{Comparing the theoretical and the observed Stokes profiles}
We compute the theoretical Stokes profiles using a procedure similar to the one described in \citet[][see also 
\citealt{2011ApJ...737...95A,2012A&A...541A..24S,2013ApJ...768..163S}]{2005A&A...434..713H}. 
We use the PRD-capable MALI (Multi-level Approximate Lambda
Iteration) code developed by \citet[][referred to as the RH-code]{2001ApJ...557..389U}
to compute the opacities, collision rates and intensity. These quantities are then given as inputs 
to the polarized radiative transfer equation defined in Equation~(1) of \cite{2013ApJ...768..163S}.
The expressions used to compute quantities such as the line and continuum source vectors, and their sum 
are same as Equations~(5), (9), and (4) of \cite{2013ApJ...768..163S}.
The Stokes profiles are computed by solving the transfer equation perturbatively 
\citep[see][]{1987A&A...178..269F,2002A&A...395..305N}.

The Sc~{\sc ii} atom model which is used as an input to the RH-code is constructed with
eight $J$-levels coupled by six line transitions 
and ten continuum transitions. The main line is treated in PRD and the 
others are treated under the assumption of complete frequency redistribution (CRD).
The angle averaged redistribution functions of \cite{1962MNRAS.125...21H} are used for intensity computations
and the angle-averaged redistribution matrix presented in \cite{2013ApJ...768..163S} is used
for computing the polarization. The continuum is computed
by assuming frequency coherent scattering and its expression is given in Equation~(9) of \cite{2013ApJ...768..163S}.
In the RH-code, all the blend lines are assumed to be depolarizing and are treated under LTE.

We compare the observed and the theoretical profiles computed using 
several model atmospheres and this is discussed below, in detail.

\begin{table} [ht]
\label{table-2}
\caption{HFS constants in MHz}
\vskip1.0pt
\centering
\begin{tabular}{ccccc}
\hline
Level & \multicolumn{2}{c}{Experiment} & \multicolumn{2}{c}{Theory} \\
\hline
\ & A & B & A & B \\
Lower & 128.2(8) & -39(11) & 146.8 & -25.5 \\
Upper & 215.7(8) & 18(7) &  202.5 & -10.8 \\
\hline 
\end{tabular}
\end{table}

\subsection{FAL models}
\label{fal-models}
\begin{figure}
\centering
\includegraphics[width=7.0cm]{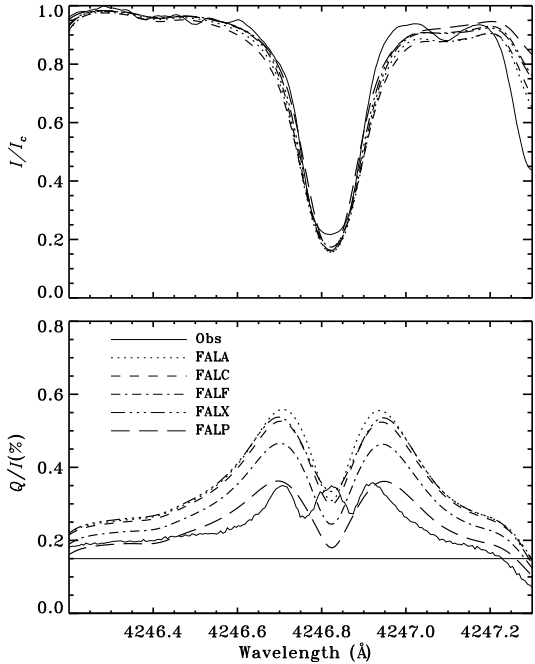}
\caption{Theoretical $(I,Q/I)$ profiles from the five standard FAL model 
atmospheres.}
\label{fig2}
\end{figure}

The FAL realistic 1D model atmospheres are taken from \citet[][FALA, FALC, FALF, and FALP models]
{1993ApJ...406..319F} and \citet[][FALX model]{1995itsa.conf..303A}. 
Figure~\ref{temp-struc}a shows their temperature structures as a function of height.
The theoretical $(I,Q/I)$ profiles computed using the FAL models are shown in Figure~{\ref{fig2}}. 
We have used the experimentally determined HFS constants given in Table~1 to 
calculate the energies of the $F$-states.
The profiles in this figure are smeared using a Gaussian with FWHM=80\,m\AA.
This smearing contains contributions from both instrument and macroturbulent velocity fields. 
Instrumental broadening is about 40\,m\AA. The rest corresponds to a macroturbulent velocity
of 2.9 km/s.

Upon comparing the theoretical and observed profiles, it is evident that 
our present treatment cannot provide an exact match to (a) the rest intensity; 
(b) the triple peak structure in $Q/I$;
and (c) the continuum polarization.

The theoretical intensity profiles are deeper than the observed profile and the 
values of rest intensity do not show much sensitivity to the variation in the model atmospheres 
within the FAL set. 
The theoretical $Q/I$ profiles do not show the triple peak structure
for any of the FAL model atmospheres whereas the observed $Q/I$ profile shows 
prominent triple peak structure.

\begin{figure}[htbp]
\centering
\includegraphics[width=7.0cm]{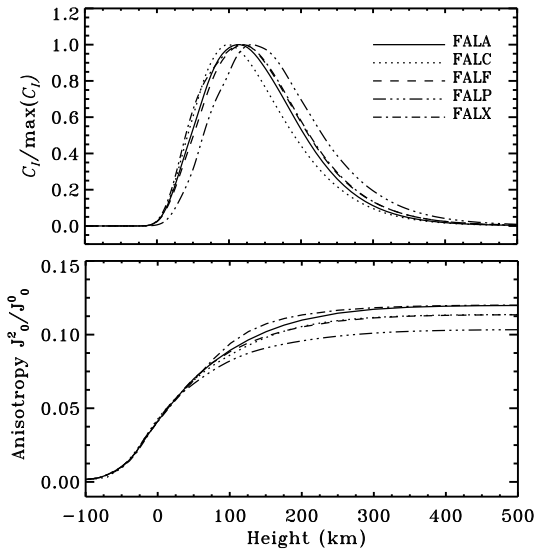}
\caption{Contribution functions and anisotropy factors 
at a continuum wavelength plotted as a function of height for several FAL model 
atmospheres.}
\label{cont-aniso}
\end{figure}

The values of the continuum polarization ($p_{c,th}$) from the FAL models 
are greater than the value determined using the
blend lines (Section~{\ref{zero-level}}). 
Such a discrepancy between $p_{c,th}$ and $p_{c,obs}$ has been studied in detail 
by \cite{2005A&A...429..713S}. In that paper, the author points out that
for $\lambda>4000$\,\AA, the $p_{c,th} > p_{c,obs}$ \citep[see Figure~6 of][]{2005A&A...429..713S}.
We also note that a similar problem with $p_{c,th}$ and $p_{c,obs}$ was 
encountered while modeling the Cr~{\sc i} triplet around 5206\,\AA\ in \cite{2012A&A...541A..24S}, and 
the model atmosphere FALF had to be modified in the deeper layers
of the atmosphere, where the continuum is formed, to fit the $p_{c,obs}$. 
Here too we face a similar problem.
Among FAL models, hotter the model atmosphere, smaller is the value of $p_{c,th}$. This is because,
as seen from Figure~\ref{temp-struc}a, the hotter models have smaller temperature gradients between 0 -- 200~km,
which leads to smaller anisotropy. The height of continuum formation can be obtained 
using the contribution functions shown in Figure~\ref{cont-aniso}. In the bottom panel of this figure,
we have plotted the anisotropy factor $J^{2}_{0}/J^{0}_{0}$ \citep[defined for instance in][]{2005A&A...434..713H}
at a continuum wavelength as a function of height. Among the FAL models, the FALP model gives the smallest
anisotropy for the continuum. This results in smaller values of $p_{c,th}$.
The values of $p_{c,th}$ computed from the FAL models are given in Table~\ref{cont-pol}.

The difficulties in reproducing the observed Stokes profiles
can originate from various sources such as an incorrect choice of model atmosphere,
multidimensional radiative transfer effects, an incomplete treatment of atomic physics, etc. 
To shed light on these issues we now try to model the observed profiles using 
another set of models with temperature structures different from the FAL models.

\subsection{FCHHT models}
The FCHHT models are the updated models published by \cite{2009ApJ...707..482F}. 
The temperature structures of these models as a function of height are shown in
Figure~\ref{temp-struc}b.
For comparison, the temperature structures of the FALC and FALP models are also shown. 
As seen from this figure, the temperature gradient in the FCHHT models are similar 
to the FAL models upto 500 km and then vary from the FAL models. Among the FCHHT models, 
the P model has the least temperature gradient which also is the hottest model. 
The B model, has the largest gradient and is also the coolest. 
In the height range 0-500 km, among the FAL models, it is the FALP model which has the least gradient,
and this gradient is smaller than the FCHHT-P model too. A comparison between the FAL and FCHHT models
is also presented in \cite{2012A&A...540A..86R}.

\begin{figure}[htbp]
\centering
\includegraphics[width=7.0cm]{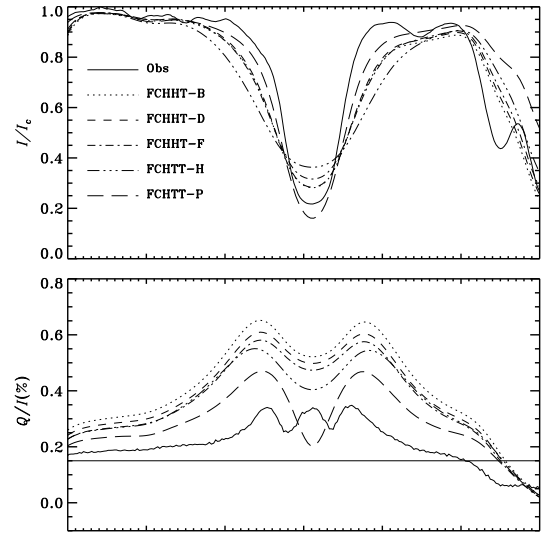}
\caption{Stokes $(I,Q/I)$ profiles of the Sc~{\sc ii} 4247\,\AA\ line computed using the 
FCHHT models.}
\label{fchht-profiles}
\end{figure}

The $(I,Q/I)$ profiles computed 
from the FCHHT models are shown in Figure~{\ref{fchht-profiles}}. We find that the updated models 
too fail to reproduce the observed Stokes profiles. We continue to face difficulties 
in reproducing the rest intensity, triple peak structure in $Q/I$, and continuum polarization. 
We discuss each of these issues in the sections below.

\subsubsection{Intensity profile}
In addition to a mismatch in line depth, the intensity profiles from the 
FCHHT models are wider than the observed profile.
The FCHHT-B, D, F, and H models produce intensity profiles wider and shallower than the 
observed profile whereas the FCHHT-P model, produces profile deeper than the observed one.
Hence in principle it should be possible to construct a two-component model, like the one 
described in \cite{2006A&A...449L..41H}, by 
mixing two appropriate models to get the required rest intensity. However, such a fit would 
give an intensity profile that is much too wide. 
There also could be a role of multidimensional radiative transfer effects
in shaping the intensity profiles. However, the scope of the present paper is restricted to 
1D models only. Therefore it is hard to quantify to what extent the multidimensional 
effects play a role.

\subsubsection{Central peak in $Q/I$ profile}
The FCHHT models have different temperature gradients in the upper 
chromospheric layers, where the line is formed, compared to the FAL models. However, this is of 
little help in improving the fit to the triple peak  structure in $Q/I$. 
Hence, the failure of both FAL and FCHHT models, with very different temperature structures, 
and the fact that the central peak is mostly governed by the $F$-state interference effects,
indicates that the mismatch at the central peak in $Q/I$ is originating from an incomplete 
treatment of atomic physics. An effect not accounted for, in our treatment, is the lower-level Hanle effect.
\cite{2009A&A...508..933B} showed that this effect can increase the 
amplitude of the central peak in $Q/I$ when the magnetic field strength, $B\approx1$G.

\subsubsection{Continuum polarization}
The $p_{c,th}$ from the FCHHT models are similar in magnitude to the ones predicted by the FAL models.
Among FCHHT models, the warmer model, FCHHT-P, predicts smaller $p_{c,th}$.
For this reason, the FCHHT-P model provides a better fit to the PRD peaks when compared to other FCHHT models.
However, this fit is not better than the one obtained from FALP model.
The values of $p_{c,th}$ from FCHHT models are given in Table~\ref{cont-pol}. Once again, none 
of these values match with $p_{c,obs}$.

The FALP and FCHHT-P models, though represent faculae  conditions, 
provide better fits to the PRD wing peaks and continuum because of the above stated reasons. 
Hence, in the sections below we conduct a few more tests using the FALP model atmosphere 
and vary some of the atomic parameters.
Note that in all the figures, the horizontal thin solid line in $Q/I$ 
represents the value of $p_{c,obs}$.

\subsubsection{Effect of macro and micro-turbulent velocities}
\begin{figure}[htbp]
\centering
\includegraphics[width=7.0cm]{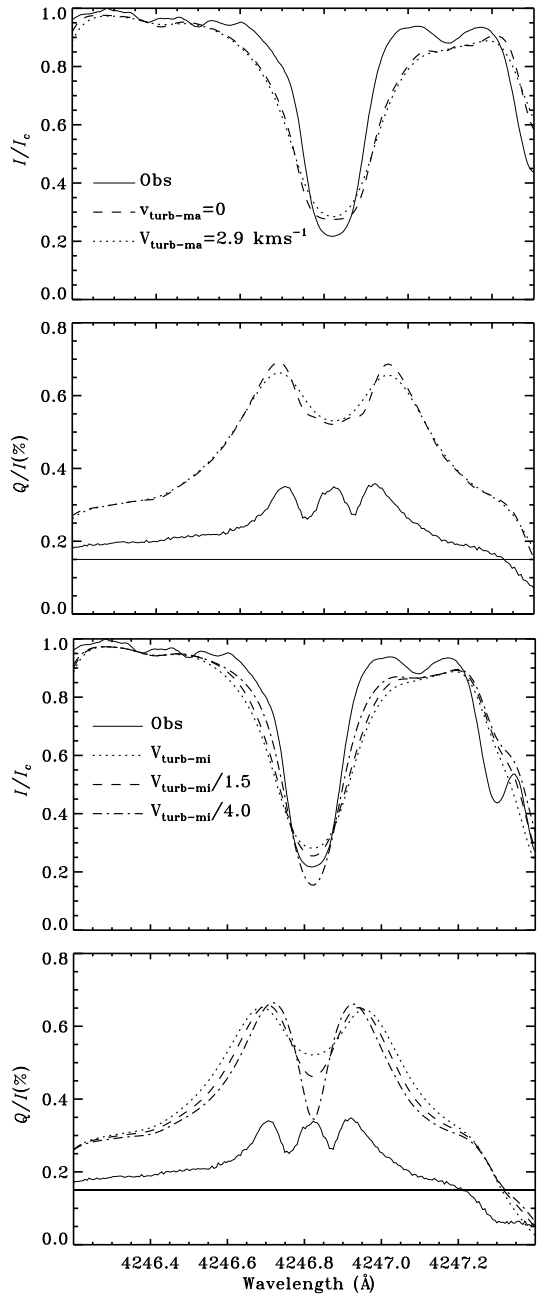}
\caption{Stokes $(I,Q/I)$ profiles of the Sc~{\sc ii} 4247\,\AA\ line computed by varying the 
macro-turbulent ($V_{\rm turb-ma}$) velocity in the top two panels and micro-turbulent velocity ($V_{\rm turb-mi}$)
in the bottom two panels. Model atmosphere used is FCHHT-B.}
\label{fchht-smear}
\end{figure}

Figure~{\ref{fchht-profiles}} shows that the intensity profiles computed 
from the FCHHT models are wide compared to the observed profile. Both macro and micro-turbulent velocities
influence the width of the intensity profiles. Reducing the macro-turbulence leads to a decrease in the smearwidth.
In the top two panels of Figure~\ref{fchht-smear}, we show a comparison between the Stokes profiles 
computed using a smearwidth of 80 m\AA\ ($V_{\rm turb-ma}$=2.9 km/s) and 40 m\AA\ ($V_{\rm turb-ma}$=0 km/s).
In the case of 4247\AA\ line we do not see a significant variation in ($I,Q/I$) with the variation in smearwidth. 
However, upon reducing the micro-turbulent 
velocity ($V_{\rm turb-mi}$) by a constant factor (1.5 and 4.0), we see that the width of the intensity
 profile decreases significantly. This is shown in 
the bottom two panels of Figure~\ref{fchht-smear}. In addition,
the central dip in $Q/I$ gets deeper. Thus it is not possible to improve the fit to 
the observed profile by varying the turbulent velocity.

\begin{table}[ht]
\caption{Continuum polarization from FAL and FCHHT models}
\label{cont-pol}
\centering
\begin{tabular}{|cccc|}
\hline
Atmosphere & $p_{c,th}$(\%) & Atmosphere & $p_{c,th}$ (\%)\\
\hline 
FALP & 0.16 & FCHHT-P  &  0.19 \\
FALF & 0.19 & FCHHT-H  & 0.22 \\
FALC & 0.21&  FCHHT-F & 0.23\\
FALA & 0.21 & FCHHT-D & 0.24 \\
FALX & 0.22 & FCHHT-B & 0.25 \\
\hline
\end{tabular}
\end{table}
\subsection{Studying the sensitivity of the $(I,Q/I)$ profiles}
We now study the response of the Stokes profiles by varying a few atomic parameters. 
We note that in some of these tests, we do recover the triple peak structure in $Q/I$.
However, the modifications made to the atomic parameters are artificial and 
are done only for the purpose of demonstrating the sensitivity of the Stokes profiles
on these parameters (see Figures~\ref{no-hfs} - \ref{hfs-mod}). 
These tests also demonstrate that the principle of spectroscopic
stability is being satisfied which proves the correctness of our treatment.
Since the intensity profiles do not show much sensitivity to these tests, we only show the 
$Q/I$ profiles.

\subsubsection{Effects of $F$-state interference}
\begin{figure}[htbp]
\centering
\includegraphics[width=6.8cm]{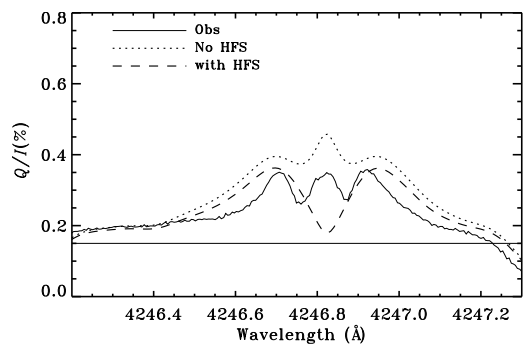}
\caption{The $Q/I$ profiles computed with and without hyperfine structure splitting.}
\label{no-hfs}
\end{figure}

It is well known that the decoherence caused by the hyperfine structure splitting of the $J$ states
leads to a depolarization in the core of the $Q/I$ line profile.
In case of the Sc~{\sc ii} 4247\,\AA\ line, the splitting between 
them is quite large and hence the decoherence.
This leads to an enhanced depolarization in the line core resulting in a
fully suppressed central peak.
When the nuclear spin $I_s=0$, we recover the triple peak structure in $Q/I$
as demanded by the principle of spectroscopic stability \citep{1994ASSL..189.....S}.
Figure~{\ref{no-hfs}} shows the comparison between the profiles with and without HFS. 

\begin{figure}[ht]
\centering
\includegraphics[width=6.8cm]{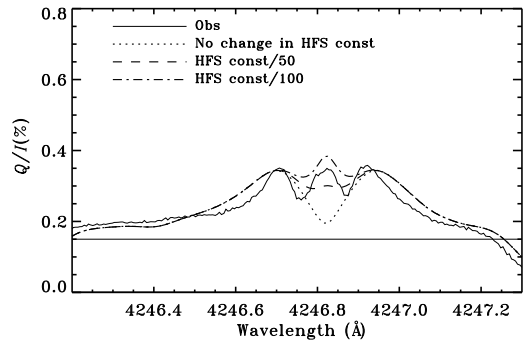}
\caption{The $Q/I$ profiles computed by reducing the HFS constants of the upper level by factors of 50 and 100.}
\label{sc-rescale}
\end{figure}

To better understand the large depolarization, we try to compare the $F$-state splitting
with the radiative widths of the upper levels. We recall that, in our treatment, the lower levels
are assumed to be infinitely sharp and hence do not interfere. 
The interfering upper $F$-states, the splitting between them and the ratio ($\Omega$) between the 
splitting and the radiative width are given in Table~\ref{omega}, where the 
Einstein $A$ co-efficient is taken as $1.29\times10^8$/s.
\begin{table*}[ht]
\caption{Comparison between the $F$-state splitting and their radiative widths} 
\label{omega}
\centering
\begin{tabular}{|ccccc|}
\hline
$F_{u1}$ & $F_{u2}$ & $\Delta E$ (Hz) & $s=(\Delta \lambda)_F$ (m\AA) & $\Omega$  \\
\hline 
3/2 & 5/2 & $5.3121\times10^{8}$  &  2.605 & 25.874\\
3/2 & 7/2 & $1.27941\times10^{9}$  & 6.274 & 62.296\\
3/2 & 9/2 & $2.24910\times10^{9}$  & 11.029 & 109.542\\
3/2 & 11/2 & $3.44605\times10^{9}$ & 16.89 & 167.844\\
5/2 & 7/2 &  $7.48200\times10^8$ & 3.669 & 36.442\\
5/2 & 9/2 &  $1.71788\times10^{9}$ & 8.424 & 83.672\\
5/2 & 11/2 & $2.91484\times10^{9}$ & 14.293& 141.972\\
7/2 & 9/2 &  $9.69685\times10^8$ & 4.755 &47.230\\
7/2 & 11/2 & $2.16664\times10^{9}$ & 10.624 & 105.518\\
9/2 & 11/2 & $1.19695\times10^{9}$ & 5.869 & 58.297\\
\hline
\end{tabular}
\end{table*}
We know that when $\Omega$ is close to unity, the splitting sensitivity is maximum.
But in case of the Sc~{\sc ii} 4247\,\AA\ line system, we see from Table~\ref{omega} that $\Omega$
is much greater than one. This partly explains the large depolarization in $Q/I$ at the line center.
{When the HFS constants are rescaled by a factor of 50 or 100, such that $\Omega$ 
approaches unity, the central peak rises up}. This again 
is a proof of the principle of spectroscopic stability being satisfied. These profiles are shown in 
Figure~\ref{sc-rescale}. Rescaling the HFS constants reduces the splitting between $F$-states and hence
the decoherence. 

\begin{figure}[ht]
\centering
\includegraphics[width=6.8cm]{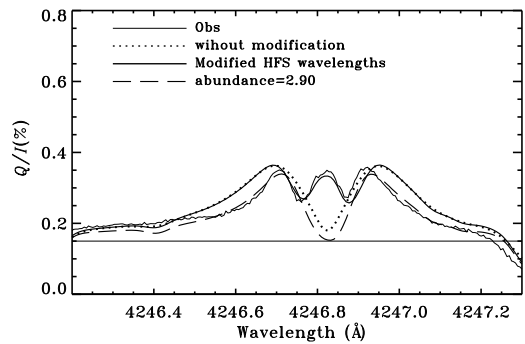}
\caption{The $Q/I$ profiles computed by modifying one of the HFS wavelengths and 
by modifying the abundance of Sc.}
\label{hfs-mod}
\end{figure}

Also, as seen from Equation~(\ref{comb-prof}) the 
$F_b=11/2 \to F_a=11/2$  is the strongest transition and it 
has maximum coupling with the $F_b=9/2 \to F_a=11/2$ transition. In other words, the 
shape of the emergent $Q/I$ profile is controlled mainly by these transitions and 
the interference between their upper levels. 
When the HFS wavelengths of these two transitions are set equal to each other, 
we recover the central peak in $Q/I$ as shown in Figure~\ref{hfs-mod}.
Such a modification, once again, reduces the decoherence and hence the depolarization. 

One can notice from Figure~\ref{hfs-mod} that the width of theoretical $Q/I$ 
profile and the amplitude of the PRD peaks are larger than in the observed profile.
Both of these are sensitive to the solar abundance of Sc.
\cite{2008A&A...481..489Z} discuss the uncertainty in the abundance value of 
Sc in the Sun. Their study is based on modeling the observed intensity
profiles of different Sc lines.
They find that different abundances are needed to fit different lines 
and conclude that the abundance value is $3.07\pm 0.04$. 
The long dashed line in Figure~\ref{hfs-mod} is the profile computed
with an abundance of 2.90. With this reduced abundance, the fit to the
PRD peaks and the near wing continuum in the $Q/I$ profile improves.

\subsubsection{Collisions}
\begin{figure}[ht]
\centering
\includegraphics[width=6.8cm]{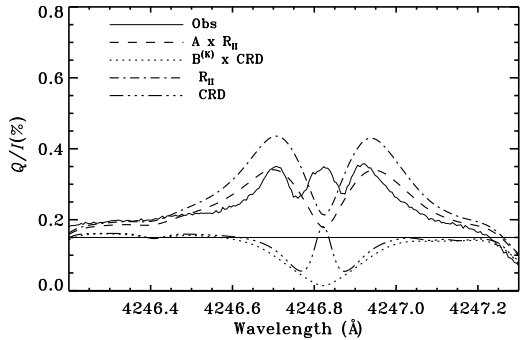}
\caption{Contributions to the $Q/I$ profile from type-II frequency 
redistribution and CRD. The profiles before and after multiplying the branching ratios are shown.}
\label{fig2a}
\end{figure}

In addition to the HFS, collisions can significantly modify the line core 
polarization of the observed profiles. 
The contribution from  collisional redistribution depends on the branching ratio $B$. 
In case of the Sc~{\sc ii} 4247\,\AA\ line, this contribution is insignificant.
Figure~\ref{fig2a} shows the individual contributions from 
type-II frequency redistribution and CRD,
with their corresponding branching ratios being multiplied. 
We note that in our computations the type-III redistribution has been 
replaced with CRD like in \cite{2012A&A...541A..24S,2013ApJ...768..163S}
to reduce the computing time. This replacement does not affect the Stokes profiles.

The variation of the branching ratios $A$ and $B^{(K)}$ as a function 
of height in the atmosphere for the 
FALP model is shown in Figure~{\ref{branch}}.  $B^{(K)}$ 
takes a value close to zero at higher layers in the atmosphere. {Since the 
Sc~{\sc ii}~4247\,\AA\ line is formed in the upper chromosphere}, the 
contribution to the line center is primarily from type-II redistribution. 
The $Q/I$ profile $B^{(K)} \times$ CRD goes nearly to zero at the line center 
(dotted line in Figure~{\ref{fig2a}}). Thus we can exclude the 
possibility that an approximate treatment of collisions 
might be contributing to the difficulties in reproducing the $Q/I$ central peak. 

Figure~\ref{fig2a} also shows the $Q/I$ profiles computed with 
type-II redistribution and CRD alone (without $A$ and $B^{(K)}$ multiplied).
The two side peaks on either side of the central peak are formed due to PRD 
and can be reproduced only by type-II redistribution. CRD alone cannot reproduce
them. Thus a proper account of PRD is essential to model this line.

\subsubsection{Variation in $\mu$}
\begin{figure}[ht]
\begin{center}
\includegraphics[width=7.0cm]{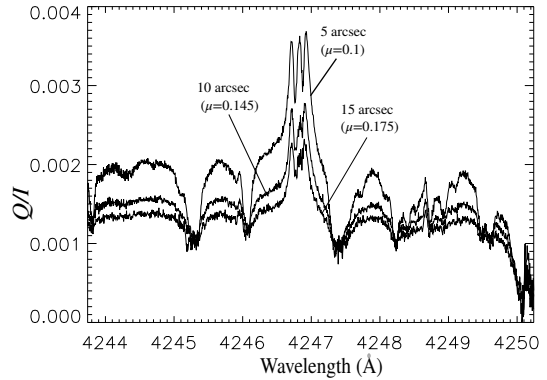}
\caption{The $Q/I$ profiles of Sc~{\sc ii} 4247\,\AA\ line observed at 
different limb distances.}
\label{allarcsec}
\end{center}
\end{figure}

\begin{figure}[ht]
\centering
\includegraphics[width=6.8cm]{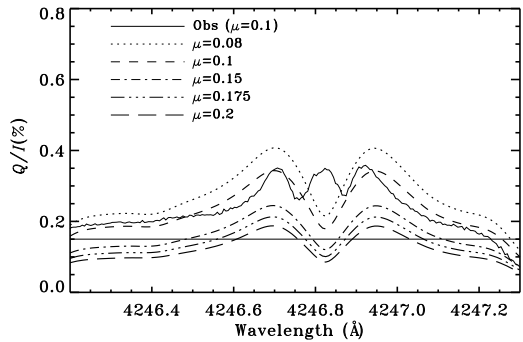}
\caption{The theoretical $Q/I$ profiles computed at various limb distances.}
\label{sc-all-mu}
\end{figure}

The observed profiles studied till now were recorded at a limb distance $\mu=0.1$.
When the line profiles were observed at nearby $\mu$ positions, they
showed a large variation in the polarization of the central peak. 
These profiles are shown in Figure~\ref{allarcsec}. At $\mu=0.145, 0.175$,  
the central peak is depolarized and only the two PRD side peaks stand out. 
The larger CLV of the central peak as compared with the side peaks 
is to be expected from spatially varying magnetic fields, since the Hanle effect 
can only operate in the Doppler line core but not in the wings.  This behavior 
is supported by the observed spatial fluctuations along the spectrograph slit: 
we find the line core amplitude of $Q/I$ to vary much more than the side peaks.
 In contrast, the theoretical profiles computed for different $\mu$ values in 
the absence of magnetic fields (cf. Figure~\ref{sc-all-mu}) do not show a variation 
of this kind.

There is therefore strong reasons to believe that the line core is greatly influenced 
by magnetic fields via the Hanle effect. This influence is normally in the form of 
depolarization, reduction of the polarization in the core. However, the Hanle effect 
may also go in the opposite direction when the atomic polarization in the lower level 
is considered, as found by \cite{2009A&A...508..933B} for fields of order 1\,G. It 
therefore remains a possibility that the observed $Q/I$ central peak that we are 
unable to reproduce with our non-magnetic modeling could be due to the Hanle effect 
in the lower atomic level.

\section{Conclusions}
In this paper, we have tried to study the Sc~{\sc ii} 4247\,\AA\ line, the polarization
profiles of which are governed by the $F$-state interference effects. 
The observations, used by us, were taken at IRSOL using ZIMPOL 3 polarimeter in September, 2012
in a quiet region near the solar limb.

Due to its large nuclear spin, the upper and lower $J$-levels 
split into five $F$-states each giving rise to thirteen radiative transitions between them.
The decoherence between the $F$-states is quite large and the emergent
polarization profiles are sensitive to the energy difference between the $F$-states. 
We have investigated 
the sensitivity of the theoretical Stokes profiles, in the absence of magnetic fields, to different 
atmospheric and atomic parameters.
All the 1-D model atmospheres tried by us (FAL and FCHHT models), fail to reproduce the triple 
peak structure in $Q/I$ and also the rest intensity. Also, the continuum polarization 
predicted by all models, except FALP, is larger than the observed value. The PRD peaks and the near wing continuum 
in the theoretical profiles match closely with the observed ones only for the FALP model, 
but this model represents faculae conditions. We also show that
a proper treatment of PRD is essential to model this line, and CRD alone cannot reproduce the PRD peaks. 

The intensity profiles computed from the FAL models match well with the observed profile
except at the line center. In case of FCHHT models, both the line width and line depth of the 
theoretical profiles match poorly with the observations. It may be possible to improve the fit to the 
intensity profiles if multidimensional radiative transfer effects are taken into account.

The $Q/I$ core peak is more sensitive to the variations in atomic parameters than the atmospheric parameters.
Observations indicate that the central peak in $Q/I$ is quite sensitive to the Hanle effect. 
There might be positive contributions from the magnetic field
to the central peak polarization through the lower level Hanle effect for field strengths of order 1\,G.

Thus, in spite of a detailed account of PRD, radiative transfer and HFS effects we are unable to 
reproduce the central peak. All these results lead us to believe that there might be other
physical effects, unaccounted for in our treatment, playing a role in shaping the $Q/I$ profiles. 
One such effect is the mentioned lower-state Hanle effect, a possibility that needs to be explored in the future.

\acknowledgments
We acknowledge the use of HYDRA 
cluster facility at the Indian Institute of Astrophysics for 
computing the results presented in the paper. 
We are grateful to  Dr. Han Uitenbroek for providing us with his 
realistic atmospheric modeling code. HNS would like to thank 
Ms. H. D. Supriya for interesting discussions. 
Research at IRSOL is financially supported by State Secretariat for
Education, Research and Innovation, SERI, Canton Ticino, the city of 
Locarno, the local municipalities,
the foundation Aldo e Cele Dacc\`o,
and the Swiss National Science
Foundation grant 200021-138016. RR acknowledges financial support by the
Carlo e Albina Cavargna foundation. We thank the Referee for 
useful comments and suggestions which helped improve the results presented in this paper.


\end{document}